\documentclass[10pt,twocolumn,table]{article}
\usepackage[top=1.9cm, bottom=1.9cm, left=1.8cm, right=1.8cm]{geometry}
\usepackage{times}  %
\usepackage[sort&compress,numbers]{natbib}  %
\usepackage[hyphens]{url}
\usepackage{graphicx}  %
\usepackage{flushend}
\usepackage[hyphens]{url}  %
\usepackage{graphicx} %
\usepackage{tikzsymbols}

\newcommand{\descr}[1]{\smallskip\noindent\textbf{#1.}}

\usepackage{sectsty}
\sectionfont{\bfseries\Large\raggedright}

\usepackage{url}                                %
\usepackage{array,multirow,graphicx,adjustbox}  %
\usepackage{booktabs}                           %
\usepackage[utf8]{inputenc}
\usepackage[ruled,algosection,noend,linesnumbered]{algorithm2e}
\usepackage{float}
\usepackage{paralist}
\usepackage{amsmath}
\usepackage{amsfonts}
\usepackage{csquotes} 
\usepackage{tabularx}
\usepackage{makecell}
\usepackage{pbox}
\usepackage[hang,flushmargin]{footmisc}
\usepackage{flushend}
\usepackage{xspace}
\usepackage{multicol, lipsum}
\usepackage[T1]{fontenc}

\usepackage{xcolor}

\usepackage{booktabs}
\usepackage{graphicx}
\usepackage{paralist}
\usepackage[bf]{caption}
\usepackage{subcaption}

\let\oldbibliography\thebibliography
\renewcommand{\thebibliography}[1]{%
  \oldbibliography{#1}%
  \setlength{\itemsep}{2pt}%
}

\usepackage[hang,flushmargin]{footmisc}

\usepackage[compact]{titlesec}
\titlespacing*{\section}{0pt}{*4}{4pt}
\titlespacing*{\subsection}{0pt}{*2.5}{2.5pt}
\usepackage{xspace}

\makeatletter
\def\url@leostyle{%
  \@ifundefined{selectfont}{\def\UrlFont{}}%
  {\def\UrlFont{}}%
}
\makeatother
\usepackage[hyphenbreaks]{breakurl}

\usepackage[bookmarks=true, bookmarksnumbered=true, colorlinks=true, linkcolor=linkcol, citecolor=citecol, urlcolor=urlcol, hypertexnames=true]{hyperref}

\definecolor{darkgreen}{RGB}{0, 100, 0}
\definecolor{linkcol}{rgb}{0.3,0,0}
\definecolor{citecol}{rgb}{0.3,0,0}
\definecolor{urlcol}{rgb}{0.3,0,0}

\makeatletter
\def\url@leostyle{%
  \@ifundefined{selectfont}{\def\UrlFont{\small}}%
  {\def\UrlFont{}}%
}
\makeatother
\newcommand{\jbnote}[1]{}
\newcommand{\synote}[1]{}
\newcommand{\edc}[1]{}

\begin{document}

\title{\bf A Data-Driven Analysis of the Sovereign Citizens Movement on Telegram\thanks{Published in the Workshop Proceedings of the 18th International AAAI Conference on Web and Social Media (ICWSM) -- Workshop: CySoc 2024: 5th International Workshop on Cyber Social Threats. Please cite the CySoc 2024 version.}}

\author{
	Satrio Yudhoatmojo,
	Utkucan Balci,
	Jeremy Blackburn \\[0.5ex]
  Binghamton University
}

\date{}

\maketitle

\begin{abstract}
Online communities of known extremist groups like the alt-right and QAnon have been well explored in past work.
However, we find that an extremist group called Sovereign Citizens is relatively unexplored despite its existence since the 1970s.
Their main belief is delegitimizing the established government with a tactic called \emph{paper terrorism}, clogging courts with pseudolegal claims.
In recent years, their activities have escalated to threats like forcefully claiming property ownership and participating in the Capitol Riot.
This paper aims to shed light on Sovereign Citizens' online activities by examining two Telegram channels, each belonging to an identified Sovereign Citizen individual.
We collect over 888K text messages and apply NLP techniques.
We find that the two channels differ in the topics they discussed, demonstrating different focuses.
Further, the two channels exhibit less toxic content compared to other extremist groups like QAnon.
Finally, we find indications of overlapping beliefs between the two channels and QAnon, suggesting a merging or complementing of beliefs.
\end{abstract}

\section{Introduction}
Online communities have brought people together and enabled many good things, like creating new forms of business and entertainment.
However, some bad things also emerged from online communities -- the spread of misinformation and the proliferation of hate speech.

Past work has explored fringe online communities that attract extremist groups to spread misinformation and disinformation, including inciting hate against other groups.
For instance, 4chan~\cite{hineKekCuckAndGodEmperorTrump2017}, Gab~\cite{zannettouWhatGabBastion2018}, and Parler~\cite{aliapouliosLargeOpenDataset2021} are well-known platforms for extremist groups like the alt-right, and the QAnon.

This paper examines an extremist group called \emph{Sovereign Citizens}.
Originating in the 1970s~\cite{Sarstechi_2020a,splc_2021}, they refute (and often outright reject) the U.S. law, delegitimizing the U.S. government, and abide by natural and common laws~\cite{Matheson_2018}.
Their well-known tactic is \emph{paper terrorism}, filling false legal documents to harass, intimidate, or extort someone, in turn clogging the courts with pseudolegal claims~\cite{Kent_2015,Sarteschi_2020b,splc_2021}.
In 2010, the FBI classified this movement as a domestic terror threat~\cite{fbi_2010}.

In recent years, Sovereign Citizens have increasingly engaged in threatening criminal behaviors, e.g., invoking arcane treaties and filing fraudulent liens to claim property ownership~\cite{davis_sovereign_2021,Nir_2021}.
Some alleged suspects for the Capitol Riot on January 6, 2021, have identified themselves as Sovereign Citizens, indicating a shift of course of action from paper terrorism towards more violent actions~\cite{Dreisbach_2021,Lybrand_2021}.

Past work about Sovereign Citizens has primarily focused on paper terrorism and their relationship with the legal system~\cite{Netolitzky_2019, Netolitzky_2021, Pytyck_2013}, leaving their activity on social media largely unexplored.
In this paper, we aim to fill the gap in the literature by investigating Sovereign Citizens' online communities.

The task of identifying this group's online community is not straightforward, unlike other extremist groups like the alt-right and the QAnon groups that past studies have shown existed on platforms like 4chan~\cite{hineKekCuckAndGodEmperorTrump2017}, Gab~\cite{zannettouWhatGabBastion2018}, and Parler~\cite{aliapouliosLargeOpenDataset2021}.
For example, on Reddit, we found subreddits like r/SovereignCitizen, r/amibeingdetained, r/SovCitCasualties, and r/MoopishConsulate, which primarily document and mock Sovereign Citizens’ antics or highlight their negative impact rather than serving as actual Sovereign Citizen communities.
On 4chan and Gab, we did not discover the Sovereign Citizens community, although 4chan's /pol/ mentioned the term ``Sovereign Citizens'' to ridicule their behavior.

Eventually, our search through article reviews~\cite{splc_2021,Kelley_2017,ADL_SovCitz_QAnon_2022,HomelandSecurityToday2022,conversation_medbeds,Sarteschi_2022} reveals two identified individuals, Bobby Lawrence and Romana Didulo, subscribing to Sovereign Citizens beliefs who have their online community on Telegram.
Telegram is a cloud-based instant messaging app that became a prominent alternative platform for extremist groups since the recent deplatforming of extremist group online communities from mainstream social media~\cite{Collier_2021,Rogers_2020,Gordon_2021,Khalil_2021,Vynck_Nakashima_2021}.
Having an identified Sovereign Citizens online community and the fact that Telegram has become an alternative platform for extremist groups, the work in this paper focuses on exploring the online activities of Sovereign Citizens who are participating in the communities of the two Sovereign Citizens' Telegram channels.

\descr{Research questions.}
Overall, we focus on four research questions:
\begin{itemize}
    \setlength{\itemindent}{2em}
    \item[\textbf{RQ1:}] Which named entities are frequently discussed in Sovereign Citizens discussions?
    \item[\textbf{RQ2:}] What topics are discussed among Sovereign Citizens in online communities?
    \item[\textbf{RQ3:}] To what extent do Sovereign Citizens exhibit toxic characteristics commonly associated with extremist groups, and how does this compare to other known extremist groups?
    \item[\textbf{RQ4:}] To what extent do Sovereign Citizens' beliefs align with other extremist groups?
\end{itemize}

\descr{Contributions.}
To the best of our knowledge, this work is the first large-scale study of Sovereign Citizens online to date.
Despite the underlying belief of delegitimizing the established government, the two Telegram channels differ in the topics they discuss.
Further, we reveal that Sovereign Citizens exhibit less toxic behavior than common extremist groups like QAnon.
Finally, our results indicate an overlapping of beliefs between Sovereign Citizens and QAnon, implying a merger or a complement of beliefs.

\section{Background and Related Work}
\subsection{The Sovereign Citizens Movement}
The Sovereign Citizens movement, originating from an anti-government Christian Identity group founded by William Potter Gale, has its roots in racism and anti-Semitism~\cite{splc_2021}. 
Referred to as the \emph{Posse Comitatus}, this group operated under the Sheriff Act of 1887, elevating the county sheriff as the supreme governmental authority~\cite{Loeser_2015, Sarstechi_2020a, splc_2021}. 
Potter's Posse engaged in illegal activities like tax evasion, property lien filing, and confrontations with authorities, which have become standard practices in the contemporary Sovereign Citizens movement. 
Over time, the movement has diversified to include members of various races, including offshoot groups focused on African Americans~\cite{southernpoveritylawcenterMoorishSovereignCitizens}.

\descr{Common Law \& Commercial Law.}
Sovereign Citizens reject the legitimacy of the U.S. government, considering it a ``corporation'' formed by corrupt bankers post the 14th Amendment ratification, leading to the rise of \emph{commercial law}~\cite{Berger2016, Sarstechi_2020a}. 
They only acknowledge the pre-14th Amendment U.S. Constitution, prioritizing \emph{common law}, which they believe overrides current U.S. laws and grants immunity from arrest and conviction. 
They see common law as God's law, unalterable by lawmakers or courts\cite{Sarstechi_2020a}. 
Additionally, they perceive indications of commercial law in court proceedings, like specific language in legal documents~\cite{Berger2016}. 
Some even go further, believing that the Uniform Commercial Code contains exploitable loopholes they can utilize when dealing with law enforcement and the legal system.

\descr{The Strawman Theory.}
Sovereign Citizens dismiss the validity of identification documents and legal contracts, such as driver's licenses, property records, and bank accounts~\cite{Berger2016}. 
They see these as tied to a fictional entity known as a \emph{strawman}, created under an illegitimate commercial law. 
To avoid interacting with the government through this strawman, some craft their own driver's licenses, license plates, and identification documents. 
They often incorporate legal phrases, citations, or unconventional punctuation in their signatures, attempting to protect their common law rights.

\descr{Declaration of Sovereignty.}
To achieve full immunity from the U.S. government, Sovereign Citizens must repudiate their strawman and assert their common law rights~\cite{Berger2016}. 
This repudiation involves a detailed process, demanding comprehension due to the specific language and legal theories involved. 
As a result, numerous individuals and groups offer legal filing kits and organize seminars to instruct on the necessary steps and actions to attain Sovereign Citizen status.

\descr{Redemption Theory.}
Sovereign Citizens claim that the U.S. government declared bankruptcy after abandoning the gold standard in 1933, resulting in foreign debt~\cite{Berger2016}.
In turn, their birth certificates or Social Security Number(SSN) are used as collateral for loans deposited into a Treasury Direct account.  
Sovereign Citizens think they can access these funds by submitting pseudolegal documents.

\subsection{Related Work}
\descr{The Sovereign Citizen Movement.}
Existing research on the Sovereign Citizens Movement, primarily from psychology and law, offers varied perspectives on the group and its members.
One of the earliest and most comprehensive descriptions of the movement comes from a Canadian divorce court ruling, Meads v. Meads~\cite{MeadsMeads2012}, where Associate Chief Justice J. D. Rooke justifies ruling against Dennis Larry Mead, who leveraged Sovereign Citizens tactics.
Spanning 140 pages, the decision covers the movement's origins, the variety of (often conflicting) theories they use, as well as examples of paper terrorism attacks like filing liens against court members, and is peppered with examples of their odd behavior (e.g., Mr. Mead filing documents signed ``:::dennis-larry:: of the meads-family::'').
Another study~\cite{Netolitzky_2019} revisits Meads v. Meads case to explore the Organized Pseudolegal Commercial Arguments (OPCA) phenomenon, considering it a valuable example for developing tools to swiftly address such litigants.

Another study~\cite{Hodge_2019} explores the roots of the movement, revealing how members disseminate their radical view of citizenship, often rooted in conspiracy theories, as a coping mechanism for anxiety.
The study of Sovereign Citizens in Australia~\cite{Baldino_2019} shows the existing Australian counterterrorism and violence framework does not cover Sovereign Citizens' activities, leading to a recommendation of new strategies that include how to disrupt their online activities and to process them.

\descr{Extremist Groups on Telegram.}
Recent moderation efforts by mainstream platforms like Facebook, Instagram, Twitter, and YouTube have reduced the overall activities of extremist groups~\cite{Walther_McCoy_2021}.
In response, these groups have migrated~\cite{Ribeiro_2021,aliUnderstandingEffectDeplatforming2021} to alternative social media platforms with lax moderation, like BitChute~\cite{trujilloWhatBitChuteCharacterizing2020,childsCharacterizingYouTubeBitChute2022}, Gab~\cite{zannettouWhatGabBastion2018}, Parler~\cite{aliapouliosLargeOpenDataset2021}, and Telegram~\cite{Alrhmoun_2023,Herasimenka_2023,Hoseini_2023,McMinimy_2023,Stewart_2023}.

Extremist Groups like the Islamic State of Iraq and Syria (ISIS) use Telegram for recruitment, radicalization, and attack coordination in Europe~\cite{Pruchan_2016,Weimann_2016,Shehabat_2017}. 
An analysis of the Islamic State's ``terrorist bot'' network on Telegram reveals its vital role in amplifying Islamic State ideology and recruiting symphatizers~\cite{Alrhmoun_2023}.
Telegram put a restriction on ISIS supporters in 2019, changing ISIS's strategy to use visual media to reinforce followers' support and concentrate their work on public agendas~\cite{McMinimy_2023}.

A study found the link between far-right actors and Telegram groups, highlighting a massive growth of far-right groups on Telegram after the 2019 deplatforming from Facebook~\cite{Urman_2020}.
A study on the QAnon communities on Telegram shows messages in Portuguese and German are more toxic than those in English, with discussions ranging from world politics to conspiracy theories and COVID-19~\cite{Hoseini_2023}.
Lastly, a study examines Australian Telegram channels linked to fascists and neo-Nazi groups and finds the channels' attempts to recruit and spread their beliefs during the COVID-19 pandemic.

\section{Dataset}
As previously noted, we decided to examine Telegram channels belonging to two identified Sovereign Citizens, Bobby Lawrence, and Romana Didulo.
Here, we present a brief introduction of who they are, followed by the data collection process and a summary of their Telegram channels.

\descr{Bobby Lawrence.}
A former U.S. Senate candidate for Pennsylvania's Republican Party in 2018~\cite{ballotpedia_bobby_lawrence_2022}, Lawrence actively encourages his followers to become American State Nationals, blending Sovereign Citizens beliefs with QAnon ideology. 
He has appeared on a QAnon talk show hosted by Ann Vandersteel, a QAnon personality. 
Lawrence spreads his teachings on his Telegram channel: \texttt{bobbylawrence\_1776}.

\descr{Romana Didulo.}
A Canadian conspiracy theorist who self-identifies as the \emph{Queen of Canada}~\cite{conversation_medbeds}.
Didulo blends Sovereign Citizens movement beliefs with QAnon conspiracy theories. 
She regularly issues decrees, like declaring free utilities and nullifying debts. 
Didulo travels across Canada in a recreational vehicle, disseminating her propaganda on her Telegram channel: \texttt{romanadidulo}.

\subsection{Data Collection}
At the time of the study, Bobby Lawrence's and Romana Didulo's Telegram channels are publicly accessible.
Lawrence's channel includes a broadcast channel and a discussion group chat, while Didulo's features broadcast and discussion channels as well as question-and-answer group chats. 
For channel and group chat definitions, refer to~\cite{Telegram,Telegram2024}.

To collect the Telegram text messages, we leverage the Telegram API~\cite{TelegramAPI_2023} within our data collection system.
This system retrieves historical messages and continuously gathers new ones. 
However, it is important to note that our data collection system cannot retrieve any messages deleted before collection begins due to API limitations.

\begin{figure}[t]
  \begin{subfigure}{.98\columnwidth}
    \centering
    \includegraphics[width=.95\linewidth]{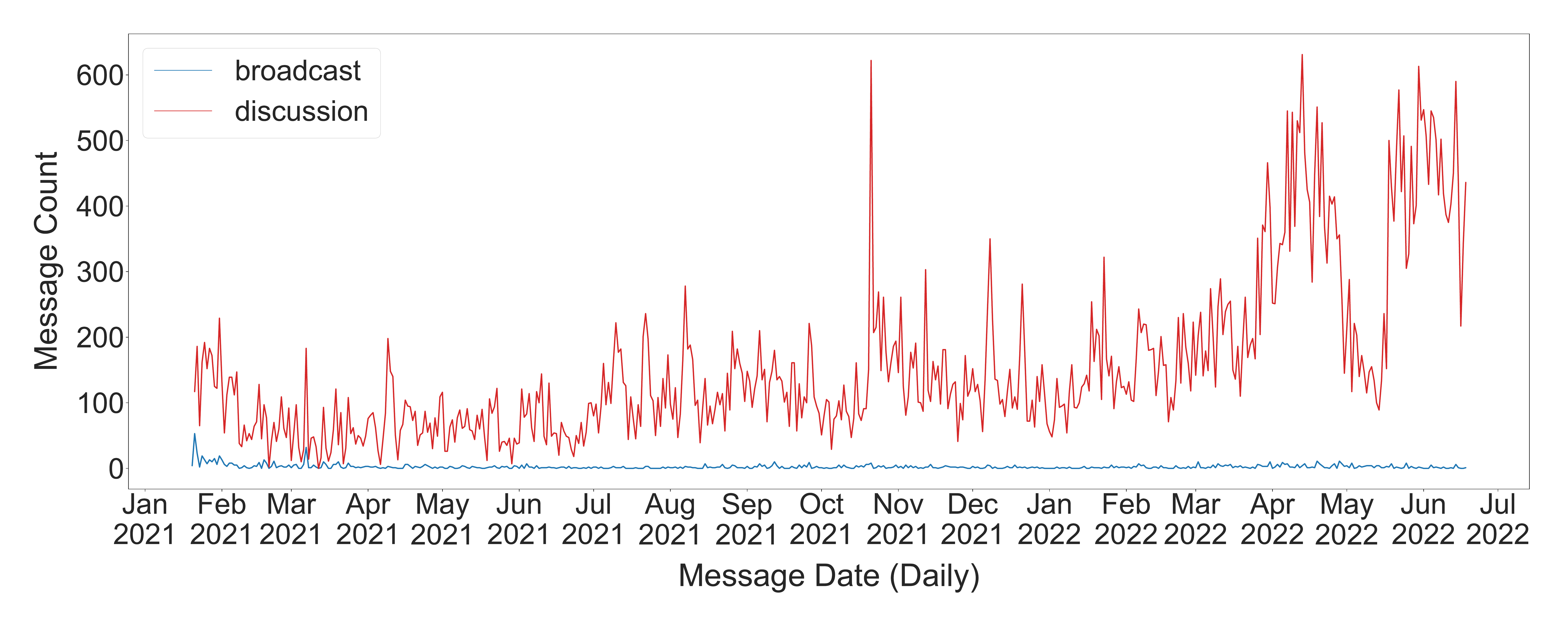}
    \caption{Bobby Lawrence's Channel}
    \label{fig:bobbylawrence_1776-message-count}
  \end{subfigure}
  \hfill
  \begin{subfigure}{.98\columnwidth}
    \centering
    \includegraphics[width=.95\linewidth]{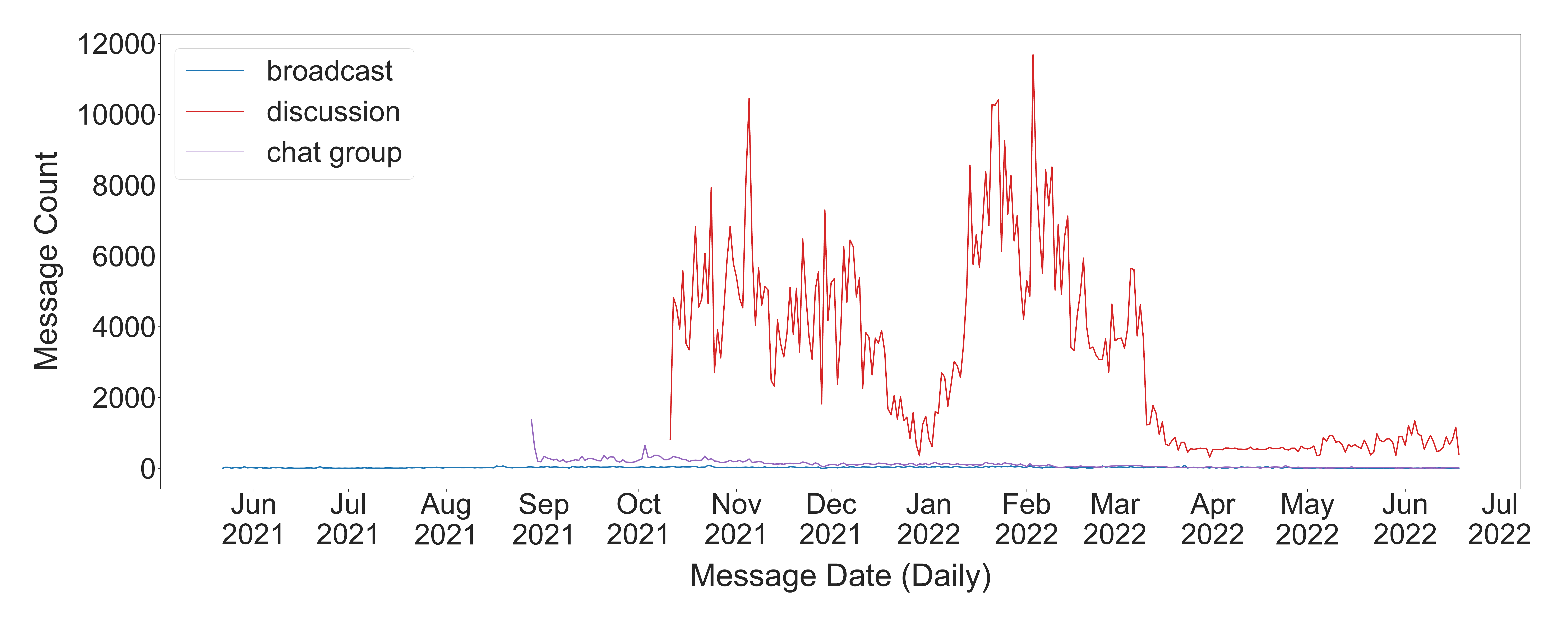}
    \caption{Romana Didulo's Channel}
    \label{fig:romanadidulo-message-count}
  \end{subfigure}
\caption{Daily message frequency. Note: the Bobby Lawrence and Romana Didulo channels have different scales for their frequency.}
\label{fig:telegram-channels-message-count}
\end{figure}

\descr{Dataset Summary.}
We collected over 888K text messages, with specifics detailed in Table~\ref{tbl:telegram-channels}.
Figure~\ref{fig:telegram-channels-message-count} shows each channel's daily message count.
This indicates that Lawrence's channel has been consistently active from the start, while Didulo's channel experienced spikes in message activity during two distinct periods.
Additionally, there are noticeable gaps in message data within Didulo's channels due to API limitations from comparing the channel creation date with the earliest dataset date in Table~\ref{tbl:telegram-channels}.
For the analysis, we combine the channel and group chat data of each Sovereign Citizen's channel.

\begin{table*}[t]
  \small
  \centering
  \begin{tabular}{lllr}
    \hline
    \textbf{Channel Name} & \textbf{Created On} & \textbf{Dataset Range}& \textbf{\# of Msgs.} \\
    \hline
    Bobby Lawrence's broadcast channel & 2021/01/19 & 2021/01/20 - 2022/06/18 & 1,286 \\
    Bobby Lawrence's discussion group chat & 2021/01/21 & 2021/01/21 - 2022/06/18 & 79,895 \\
    Romana Didulo's broadcast channel & 2021/02/09 & 2021/05/22 - 2022/06/18 & 9,826 \\
    Romana Didulo's discussion group chat & 2021/01/12 & 2021/10/11 - 2022/06/18 & 762,954 \\
    Romana Didulo's Q\&A group chat & 2021/08/27 & 2021/08/28 - 2022/06/18 & 34,553 \\
    \hline
    \multicolumn{3}{r}{\textbf{Total}} & \textbf{888,514}\\
    \hline
  \end{tabular}
  \caption{List of Telegram channels, range of dates of the collected data and the number of messages.}
  \label{tbl:telegram-channels}
\end{table*}

\subsection{Ethics Statement}
We follow ethical guidelines for digital researchers~\cite{Rivers_2014}, prioritizing user privacy. 
Our findings withheld Personal Identifiable Information (PII), like usernames and phone numbers, to preserve user anonymity. 
We paraphrase quoted messages in the paper to maintain context while protecting the sender’s privacy. 
Our data collection methods, utilizing publicly accessible Telegram channels and the Telegram API, adhere to Telegram's terms of service.

\section{Results}
\subsection{RQ1: Named Entity Analysis}
\begin{figure}[t!]
  \begin{subfigure}{.98\linewidth}
    \centering
    \includegraphics[width=.95\linewidth]{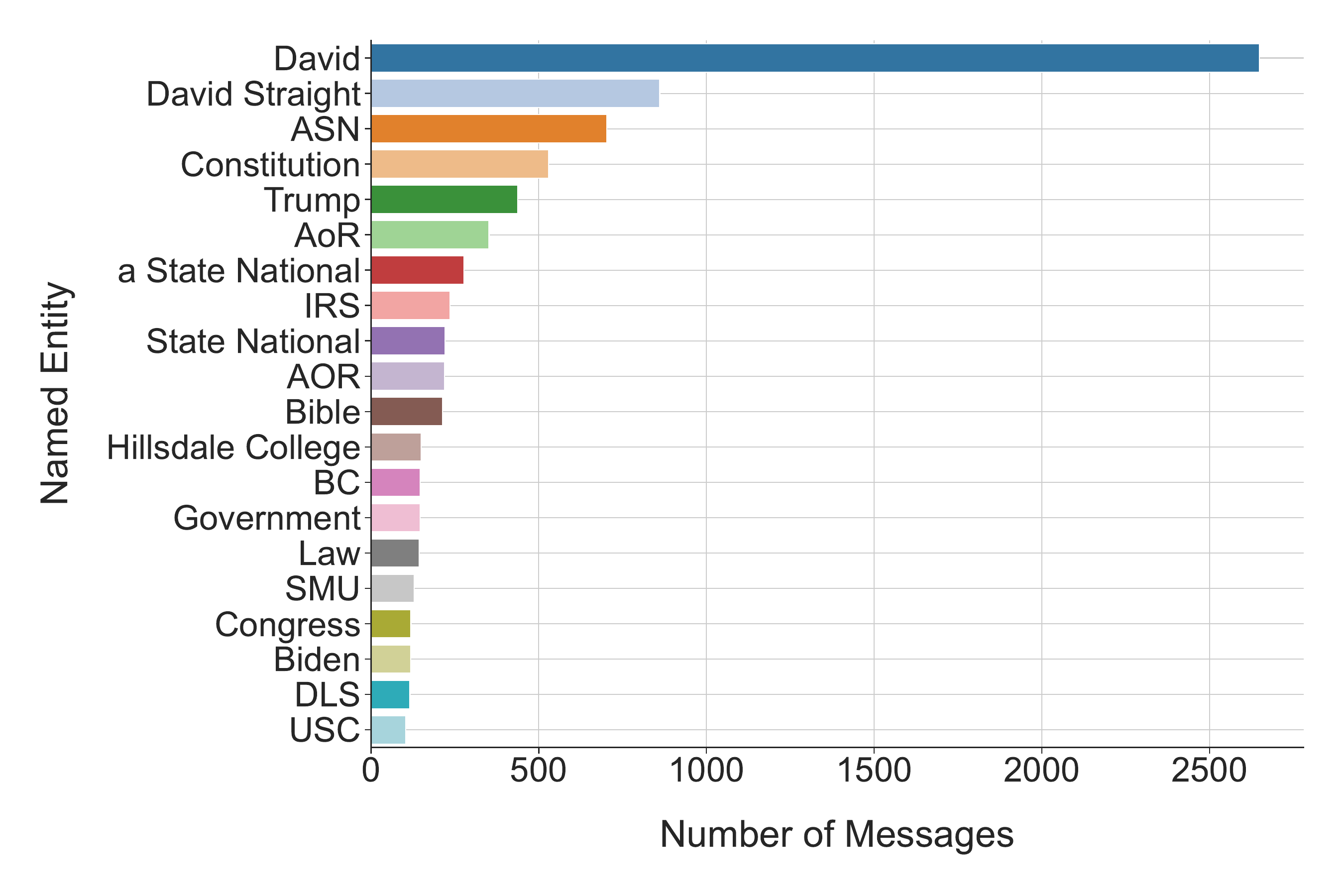}
    \caption{Bobby Lawrence's Telegram Channel}
    \label{fig:bobbylawrence_1776-ner}
  \end{subfigure}
  \vspace{2em}
  \begin{subfigure}{.98\linewidth}
    \centering
    \includegraphics[width=.95\linewidth]{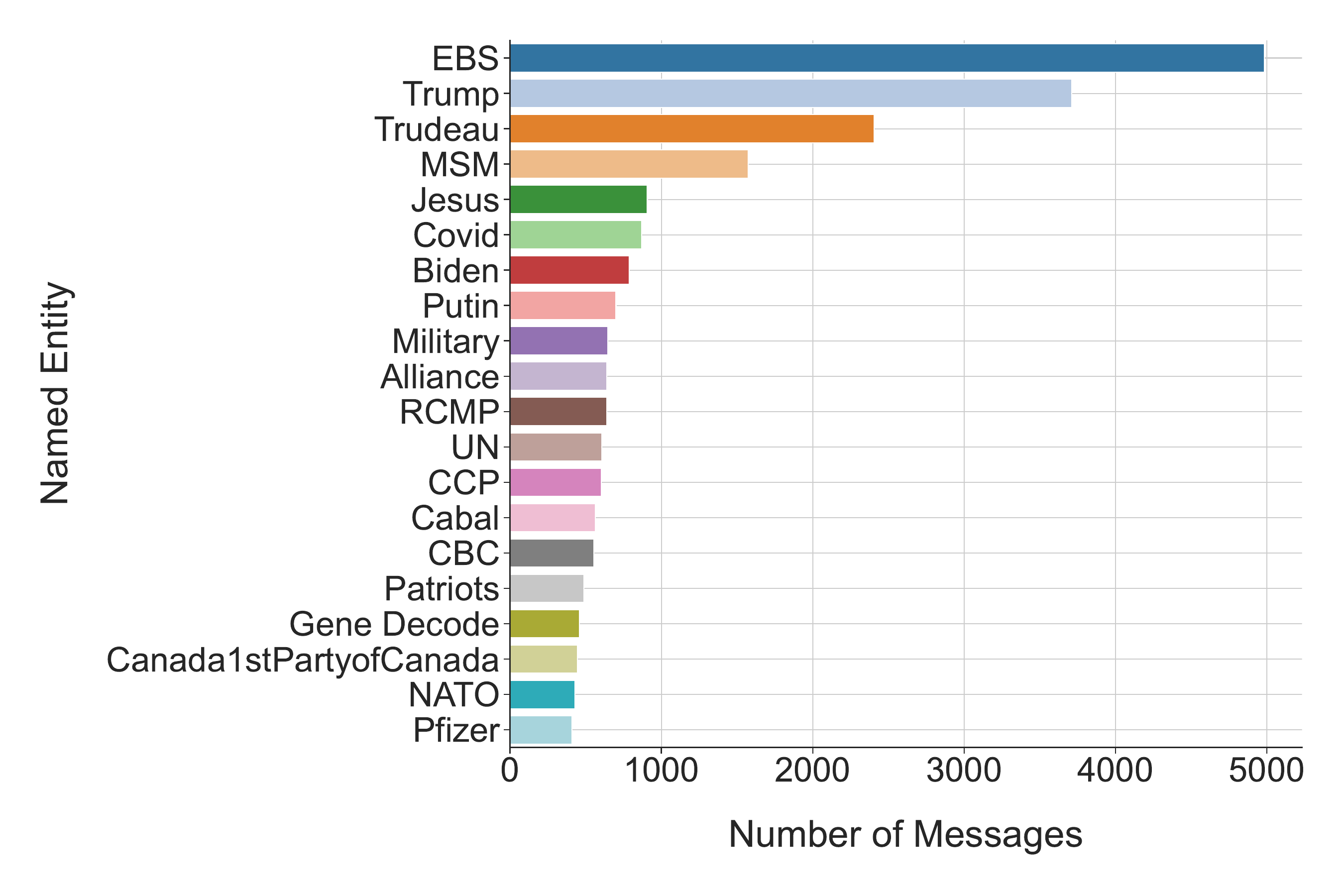}
    \caption{Romana Didulo's Telegram Channel}
    \label{fig:romanadidulo-ner}
  \end{subfigure}
  \caption{The top 20 named entities frequently discussed in the channel's messages.}
  \label{fig:top-20-ner}
\end{figure}

We leverage using Stanza~\cite{Qi_2020} to reveal named entities frequently discussed on each channel, an approach inspired by~\cite{Johansson_2016}.
All messages are combined together (i.e., broadcast and discussion messages in Lawrence's, and broadcast, discussion, and group chat messages in Didulo's).
We concentrate on four named entity types: 1)~\texttt{\textbf{PERSON}}, 2)~\texttt{\textbf{ORG}}, 3)~\texttt{\textbf{LAW}}, and 4)~\texttt{\textbf{WORK\_OF\_ART}} from the NER task result.
We remove irrelevant entities like ``Today'' and ``One'' which do not pose meaning to the context of the study.
Figure~\ref{fig:top-20-ner} shows each channel's top 20 named entities discussed.

\descr{Bobby Lawrence's channel.}
In Lawrence's channel, the named entity that occurred the most is David Lester Straight, a Sovereign Citizens guru, often referred to as ``David,'' ``David Straight,'' or ``DLS.''
Our observation of the messages mentioning entities related to David Lester Straight shows that Lawrence's channel closely follows the teaching method used by Straight, particularly about correcting a person's U.S. National status to become an American State National (ASN).
In the results, there are named entities referring to ASN: ``ASN'', ``a State National'' and ``State National.''
ASN is a strategy aimed at disassociating individuals from their strawman identity and the perceived ``corporate'' U.S. government~\cite{ADL_SovCitz_QAnon_2022,ADL_ASN_2023}.
A named entity associated with ASN is ``AoR'' or ``AOR,'' which stands for Affidavit of Repudiation, a document used to repudiate a person's U.S. National status in favor of becoming an ASN~\cite{ADL_ASN_2023}.

The channel also frequently mentions named entities like ``IRS'' and ``BC'' (Birth Certificate) which are related to the Redemption Theory~\cite{FCAS2011,Berger2016}.
Sovereign Citizens often file legitimate IRS forms with the belief that doing so will compel the U.S. Treasury to settle various debts.
Moreover, Sovereign Citizens hold the belief that birth certificates function as a corporate shell (i.e., the strawman) and collateral for loans deposited into a Treasury Direct account.

Furthermore, the channel frequently mentions named entities like ``Constitution'' and ``Congress'' which are associated with its objective to restore the pre-14th Amendment U.S. Constitution (i.e., 1787 U.S. Constitution). 
Additionally, the channel mentions ``Hillsdale College''~\cite{HillsdaleCollege2024}, which refers to the Constitution 101 YouTube video series -- a primary educational resource endorsed within the channel.

The ``Bible'' appeared in the most frequently mentioned named entities, which highlights the channel's identity as a Christian-oriented group. 
The channel believes the Constitution finds its roots in biblical teachings and that the Founding Fathers established the United States on Christian values.
Notably, our results reveal a preference for two specific versions of the Bible: the 1611 King James V Bible and the 1599 Geneva Bible.

The remaining named entities are related to government institutions, including terms like ``Government,'' ``Law,'' and ``USC'' (U.S. Code), which reflects an interest in legal frameworks and governance. 
Additionally, there is notable attention given to two U.S. presidents: ``Trump'' and ``Biden.'' 
Another noteworthy named entity is ``SMU,'' an abbreviation for ``Shit Made Up,'' which the channel uses to dismiss government entities or concepts considered illegitimate by the group.

\descr{Romana Didulo's channel.}
In contrast to Lawrence's channel, Didulo's channel engages with different named entities. 
The most frequently named entity that the channel engages with is ``EBS''(Emergency Broadcast System), where Didulo promises to alert the world about military actions to apprehend wrongdoers and initiate a series of disclosures lasting eight hours, introducing a new financial system~\cite{Nebula2021}. 
However, this promise remains unfulfilled, leaving followers speculating about the timing of the EBS announcement.

Additionally, the channel shows fixations on various world leaders like ``Trump,'' ``Trudeau,'' ``Putin,'' and ``Biden.'' 
Analysis of the messages indicates support for Trump and Putin, with quotes expressing gratitude for their actions:
\begin{quote}
    \small
    \emph{``Trump was chosen by Romana and deserves our love for his great sacrifice and service to We the People.''}
\end{quote}
\begin{quote}
    \small
    \emph{``Romana has several times told us Putin is good.''}
\end{quote} 
\noindent Conversely, there is negativity toward Trudeau and Biden, as evidenced by quotes criticizing their actions:
\begin{quote}
    \small
    \emph{``Fake Trudeau, the puppet, is an a**-kisser who follows the same cabal agenda as fake Biden''}
\end{quote} 
\noindent The fixation on ``Cabal,'' which is also evident in discussions about world leaders, suggests adherence to the New World Order conspiracy theory~\cite{American_Jewish_Committee_2023d}. 
This ties into fixations on government entities like the ``UN,'' ``NATO,'' and ``CCP.''

Moreover, the channel expresses discontent with mainstream media (``MSM'') and the Canadian Broadcasting Corporation (``CBC''), accusing them of spreading false narratives. 
Additionally, there are fixations on ``COVID'' and ``Pfizer,'' reflecting beliefs in COVID-19 conspiracy theories. 
For instance, one message suggests ongoing propaganda surrounding COVID-19 and the promotion of vaccines:
\begin{quote}
    \small
    \emph{``It’s still happeing; they are still doing the propaganda about Covid-19, nothing has changed, they still are promoting vaccines/ poison schedules.''}
\end{quote}

\descr{Key Takeaways.}
The frequently named entities in each channel shed light on their respective agendas. 
Bobby Lawrence's channel focuses on agendas like American State National and restoring the pre-14th Amendment U.S. Constitution. 
The top 20 named entities in Lawrence's channel closely align with various Sovereign Citizens' beliefs and theories. 
Conversely, Romana Didulo's channel often references global events like the Russian invasions of Ukraine and COVID-19. 
However, message analysis reveals a prevailing distrust towards established governments, including the Canadian government, relevant to Sovereign Citizens' beliefs.

\subsection{RQ2: Topic Analysis}
In the previous analysis, we uncovered named entities mentioned in each channel.
Despite indicating what the channel is focusing on, we did not analyze the discussion topics on the channel.
To better understand the topics discussed in each channel, we perform a topic modeling task using Top2Vec~\cite{Angelov_2020}, a topic modeling algorithm suitable for analyzing topics from short messages.
An advantage of Top2Vec is that there is no requirement to preprocess the texts to generate the topic model.
However, it can result in too many, overly fine-grained topics.
Therefore, we apply topic reduction using Top2Vec's \texttt{hierarchical\_topic\_reduction} method.
We reduce the topic to a maximum of 20, allowing us to qualitatively analyze the generated topics.
Additionally, we use ChatGPT to summarize each topic's representative documents (i.e., the Telegram messages), allowing us to speed up our analysis.
We run the following simple ChatGPT prompt: \emph{``Summarize the following Telegram messages: [a list of Telegram messages in the topic documents]''} to summarize the topic documents.

\begin{table}[t!]
  \small
  \begin{subtable}{0.98\columnwidth}
    \centering
    \begin{tabular}{>{\centering\arraybackslash}p{0.1\columnwidth}p{0.8\columnwidth}}
        \hline
        \textbf{Topic \#} & \textbf{Topic Description} \\
        \hline
        1 & American State National Status Correction Process. \\
        2 & Restoring the 1787 U.S. Constitution. \\
        3 & Request for Seminar Schedule by Location. \\
        4 & Restoration of Constitutional Republic. \\
        5 & Lawful vs. Legal Concepts. \\
        6 & Knowledge Building: Channel Content Exploration. \\
        7 & State Nationals vs. U.S. Nationals. \\
        8 & Homeschooling as Education Preference. \\
        9 & Financial Complexity and Legal Concepts. \\
        10 & Challenges in Obtaining Birth Certificates. \\
        \hline
    \end{tabular}
    \caption{Bobby Lawrence's channel.}
    \label{tab:bobby-lawrence-topics}
  \end{subtable}
  \hfill
  \begin{subtable}{0.98\columnwidth}
    \centering
    \begin{tabular}{>{\centering\arraybackslash}p{0.1\columnwidth}p{0.8\columnwidth}}
        \hline
        \textbf{Topic \#} & \textbf{Topic Description} \\
        \hline
        1 & Current Affairs: Global Conflicts and Govt. Distrust. \\
        2 & 10D Vibrational Videos Recommendation. \\
        3 & COVID-19 Controversies. \\
        4 & Questioning Religious Mainstream Narratives. \\
        5 & Freedom Convoy. \\
        6 & Royal Decree about Tax. \\
        7 & Support on Freedom Convoy. \\
        \hline
    \end{tabular}
    \caption{Romana Didulo's channel.}
    \label{tab:romana-didulo-topics}
  \end{subtable}
  \caption{List of discussion topics generated using Top2Vec.}
  \label{tab:topic-model}
\end{table}

\descr{Topics in Bobby Lawrence's Channel.}
The topic modeling task generates 631 topics, which is reduced to 20 topics.
After analyzing the topic documents with assistance from ChatGPT, the results reveal ten relevant topics (see Table~\ref{tab:bobby-lawrence-topics}).
Overall, the discussion topics revolve around citizenship (e.g., American State National), the Constitutional Republic, legal education, birth certificates, homeschooling, and Bobby Lawrence's seminar.

In Topic \#1, we reveal a key discussion revolving around the status correction process to become an American State National.
One of the messages mentioned a \emph{Freedom Package}, a guide to status correction.
Related to this, Topic \#7 emphasizes the difference between State Nationals and U.S. Nationals, stating that they do not renounce their citizenship but rather repudiate it from U.S. Nationals to State Nationals.

Other topics revolve around the Constitution (Topics \#2 and \#4).
These topics encourage the followers to learn about U.S. history to restore the nation to the 1787 U.S. Constitution, which they consider the ``true'' U.S. Constitution.
Related to this, Topic \#8 reveals the channel's preference of homeschooling their children, believing that the educational system is flawed and not teaching the children the ``true history'' of the United States.

We also see topics that discuss legal concepts.
Topic \#5 focuses on the difference between lawful and legal concepts.
The former suggests it is something based on God's law, natural law, or common law, while the latter is based on the laws of man or man's construct.
Following this topic is Topic \#9, which distinguishes between legal tender (related to Federal Reserve Notes) and lawful money (related to gold and silver coins).
The topic also mentioned CQV trust, which, according to Redemption Theory, is related to funds obtained by the U.S. government from loaning the person's birth certificate.

\descr{Topics in Romana Didulo's Channel.}
The topic modeling task initially generated 3,692 topics, which we reduced to 20. 
Among these, only seven are relevant. 
These topics, detailed in Table~\ref{tab:romana-didulo-topics}, are entirely different from those found in Bobby Lawrence's channel, centering on four main themes: current affairs, COVID-19, skepticism towards mainstream narratives, and Romana Didulo's antics.

Topic \#1 explores various current events, offering unique perspectives. 
For instance, during the 2022 Russian invasion of Ukraine, the discussion shifts towards beliefs that Russia is aiding Ukraine to dismantle bioweapon labs and confront Ukrainian Nazis. 
Additionally, the channel uses terms like ``corrupt corporation,'' typical of Sovereign Citizen jargon, and ``white hat,'' a term often associated with Trump supporters in QAnon conspiracy theory.

Other important topics include \#3, \#5, and \#7, which center around COVID-19 discussions. 
Messages within these topics reflect skepticism towards the origins of COVID-19, believing the COVID-19 pandemic was planned, and spreading misinformation about a fictitious entity, the International Common Law Court of Justice, falsely convicting pharmaceutical companies and government officials of genocide. 
Moreover, these topics highlight support for the Freedom Convoy protest against COVID-19 mandates, especially with the participation of Romana Didulo in the protest.

Topic \#4 continues the channel's divergence from mainstream narratives by scrutinizing religious institutions. 
The discussion here challenges traditional beliefs about the Bible and the Vatican, suggesting that the latter uses fear to control Christians and potentially conceals additional biblical texts.

Lastly, Topics \#2 and \#6 shed light on Romana Didulo's antics. 
Topic \#2 focuses on videos recommended by Didulo, which are claimed to be infused with unique vibrational frequencies. 
These videos cover diverse topics such as health, spirituality, and energy clearing. 
Meanwhile, Topic \#6 explores the royal decrees issued by Didulo as the self-proclaimed Queen of Canada, particularly focusing on tax-related decrees, which some followers struggled to implement despite claims of Canada 2.0's tax-free status and the non-existence of the Canada Revenue Agency (CRA).

\descr{Key Takeaways.}
Lawrence's channel  focuses on educating followers about becoming American State Nationals and restoring what they see as the true U.S. Constitution. 
There is a strong emphasis on homeschooling due to distrust in the mainstream U.S. history curriculum. 
In contrast, Didulo's channel explores unconventional perspectives on current events and discusses Romana Didulo's royal decrees. 
However, these decrees lack real-world applicability, leading to confusion among followers.

\subsection{RQ3: Toxicity Analysis}
It is widely recognized that content in extremist groups' communities is highly toxic~\cite{Papasavva_2021,Hoseini_2023}. 
Therefore, we measure the toxicity of content within Sovereign Citizens' communities. 
To do this, we use the Perspective API~\cite{Perspective} to measure the content toxicity of Bobby Lawrence's and Romana Didulo's channels, focusing on the API's SEVERE TOXICITY attribute scores. 

Additionally, we introduce three new datasets for comparison: a 0.5\% random sample of Reddit posts and Telegram messages created between January 20, 2021, and June 18, 2022, curated by Pushshift~\cite{Baumgartner2020,Baumgartner2020a}, and a QAnon Telegram dataset from~\cite{Hoseini_2023}, consisting of messages created from September 17, 2021, to March 9, 2021. 
All three datasets' content toxicity is also measured using the Perspective API.
The 0.5\% random sample of Reddit and Telegram data serve as arbitrary, generally non-specific online posts to compare with our Sovereign Citizens datasets.
While the QAnon Telegram dataset represents the same group category (i.e., extremist group) which shows high numbers of toxic content~\cite{Hoseini_2023}.
Therefore, the three datasets provide a balanced comparison to our datasets.

\begin{table}[t]
  \small
  \centering
  \begin{tabular}{lrrrrr}
    \hline
    \textbf{Dataset} & \textbf{Msgs.} & \textbf{Max.} & \textbf{Median} & \textbf{\% $>$ 0.8}\\
    \hline
    Bobby Lawrence  & 75K   & 0.46    & 0.0022 & 0\% \\
    Romana Didulo   & 513K  & 0.69    & 0.0024 & 0\% \\
    Reddit 0.5\%    & 19M   & 1       & 0.0021 & 0.05\% \\
    Telegram 0.5\%  & 604K  & 0.99    & 0.0017 & 0.26\% \\
    QAnon Telegram  &  2M   & 1       & 0.0405 & 1.8\% \\
    \hline
  \end{tabular}
  \caption{The severe toxicity scores for the four datasets. It shows the total number of messages scored by Perspective API followed by the maximum, and median scores of Perspective API's $\texttt{SEVERE\_TOXICITY}$ attribute, and the percentage of messages scored above 0.8.}
  \label{tbl:toxicity-scores}
\end{table}

\begin{figure}[t]
  \centering
  \includegraphics[width=.98\linewidth]{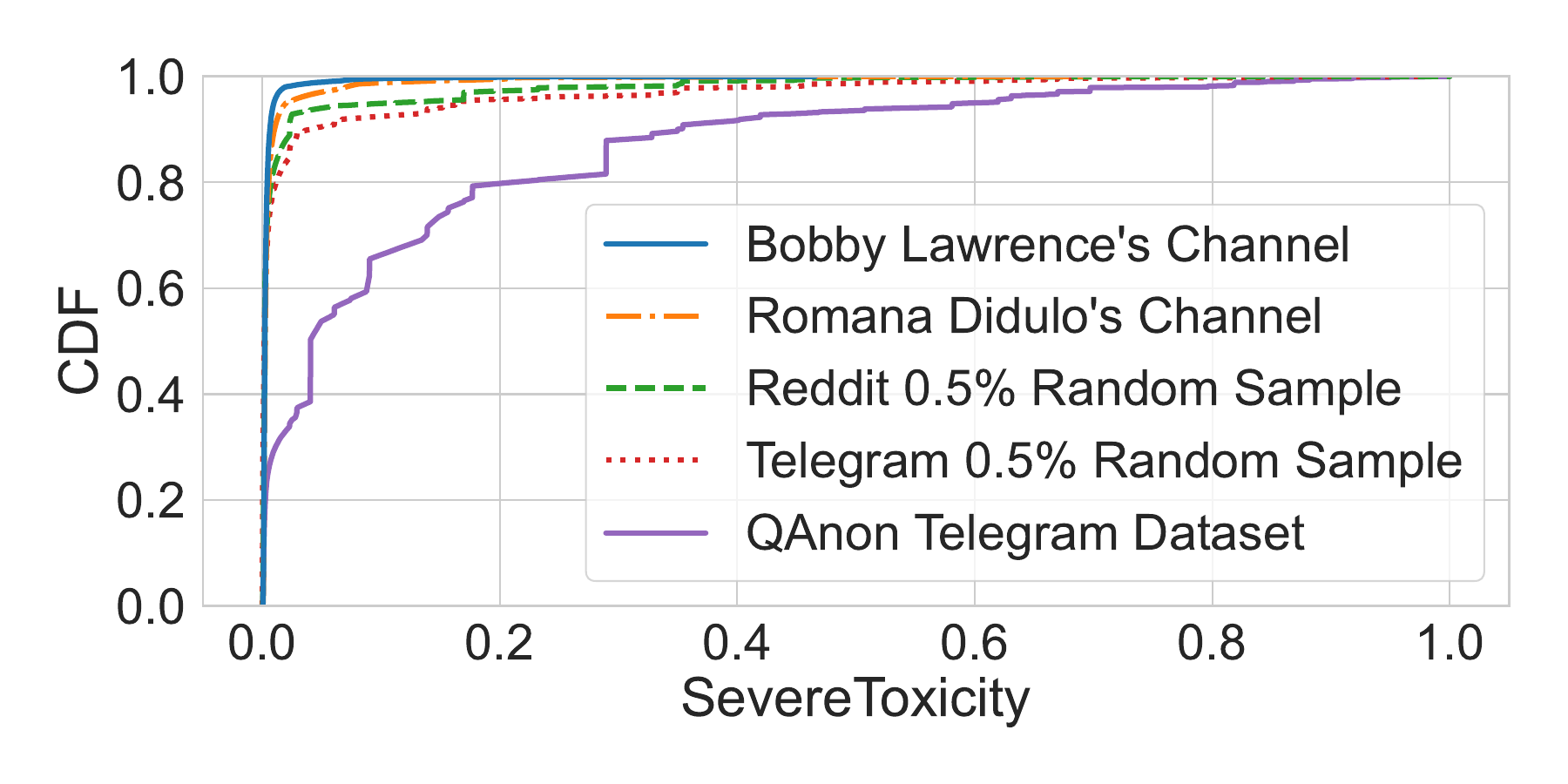}
  \caption{The CDF of severe toxicity scores.}
  \label{fig:severe-toxicity-scores}
\end{figure}

We do not preprocess the text in our Sovereign Citizens datasets and the three comparing datasets in measuring the toxicity score using Perspective API.
To determine the minimum threshold of a message considered severely toxic, we follow~\cite{Hoseini_2023} that uses the toxicity score $> 0.8$ as a threshold.
Table~\ref{tbl:toxicity-scores} presents the number of messages scored, the toxicity score statistics (minimum, maximum, median, and mean score values), and the percentage of messages with scores above 0.8 for each dataset.
Figure~\ref{fig:severe-toxicity-scores} shows the cumulative distribution functions (CDFs) of the toxicity scores of the four datasets.

The results show that content from Bobby Lawrence's and Romana Didulo's channels exhibit the lowest toxicity, with maximum severe toxicity scores of 0.46 and 0.69, respectively -- none of the messages pass the severe toxicity threshold.
These findings indicate that the Sovereign Citizens' channels do not use toxic language.
In contrast to Sovereign Citizens, the QAnon Telegram dataset is the most toxic dataset with a median of 0.0405, and 1.8\% of messages are severely toxic. 
The 0.5\% random sample of Reddit and Telegram datasets are between the Sovereign Citizens' channels and the QAnon Telegram datasets, having both non-toxic text content and severely toxic content.
Additionally, we run a 2-sample Kolmogorov-Smirnov test to check for statistically significant differences between all the distributions in Figure~\ref{fig:severe-toxicity-scores} and find the p-value on each pair less than 0.01 ($p < 0.01$).

\descr{Key Takeaways.}
Our findings indicate that the content from Bobby Lawrence's and Romana Didulo's channels does not exhibit toxicity, scoring 0.0022 and 0.0024 as their median scores, respectively. 
In comparison, the QAnon Telegram dataset shows a higher level of toxicity than the Sovereign Citizens dataset, with a median score of 0.0405. 
This contrast highlights that while Sovereign Citizens may employ legal tactics perceived as toxic in the real world, their online content does not reflect the same level of toxicity.

\subsection{RQ4: Alignment of Beliefs}
In this last analysis, we investigate how closely the beliefs of Sovereign Citizens align with those of another extremist group, QAnon. 
Using the QAnon Telegram dataset from our previous analysis, we employ a worldview alignment framework~\cite{Milbauer_2021}.

This framework uses shared words between the QAnon Telegram dataset and each of our Sovereign Citizens Telegram channels to perform a linear transformation using MultiCCA~\cite{ammar2016massively} to project all shared words from each of our Sovereign Citizens datasets to the QAnon Telegram dataset.
The framework uses cosine similarity to measure how close the linear transformation result is.
If the nearest image of the word is the word itself, then we say that it has reached a perfect alignment.
Otherwise, they are misaligned.
Another output of this framework is a pair of conceptual homomorphism words that show two similar worldviews/beliefs and conceptual structures but talk about different things.
We use the same preprocessing and training procedure suggested by~\cite{Milbauer_2021}, including using the top  5,000 words from the shared vocabulary of each community pair as ``anchor words,'' as it provides similar accuracy compared to using all shared vocabulary.
We include common trigrams in addition to common bigrams.

\begin{table}[t!]
  \small
  \centering
  \begin{subtable}[t!]{0.49\textwidth}
    \centering
    \begin{tabular}{llr}
      \hline
      \textbf{Bobby Lawrence} & \textbf{QAnon} & \textbf{Alignment} \\
      \hline
      \cellcolor{red!30}natural law & \cellcolor{red!30}common law & \cellcolor{red!30}0.4718 \\
      \cellcolor{yellow!30}illuminati & \cellcolor{yellow!30}hillary clinton & \cellcolor{yellow!30}0.4318 \\
      \cellcolor{red!30}corporations & \cellcolor{red!30}government & \cellcolor{red!30}0.4308 \\
      \cellcolor{green!30}fauci & \cellcolor{green!30}bill gates & \cellcolor{green!30}0.4194 \\
      \cellcolor{red!30}fiat & \cellcolor{red!30}cryptocurrency & \cellcolor{red!30}0.4140 \\
      \cellcolor{green!30}vax & \cellcolor{green!30}covid & \cellcolor{green!30}0.4139 \\
      \cellcolor{yellow!30}deep state & \cellcolor{yellow!30}globalist & \cellcolor{yellow!30}0.4090 \\
      \cellcolor{yellow!30}ds & \cellcolor{yellow!30}alien invasion & \cellcolor{yellow!30}0.3921 \\
      \cellcolor{green!30}jab & \cellcolor{green!30}vaccines & \cellcolor{green!30}0.3858 \\
      \hline
    \end{tabular}
    \caption{Bobby Lawrence channel -- QAnon Telegram dataset.}
    \label{tbl:bobbylawrence-1776-alignment}
  \end{subtable}

  \hfill

  \begin{subtable}[t!]{0.49\textwidth}
    \centering
    \begin{tabular}{llr}
      \hline
      \textbf{Romana Didulo} & \textbf{QAnon} & \textbf{Alignment}  \\
      \hline
      \cellcolor{green!30}protesters & \cellcolor{green!30}rioters & \cellcolor{green!30}0.6285 \\
      \cellcolor{yellow!30}cabal & \cellcolor{yellow!30}globalist & \cellcolor{yellow!30}0.6282 \\
      \cellcolor{red!30}corporations & \cellcolor{red!30}governments & \cellcolor{red!30}0.5760 \\
      \cellcolor{red!30}bankrupt & \cellcolor{red!30}corporation & \cellcolor{red!30}0.5529 \\
      \cellcolor{green!30}scamdemic & \cellcolor{green!30}pandemic & \cellcolor{green!30}0.5521 \\
      \cellcolor{yellow!30}deep state & \cellcolor{yellow!30}communists & \cellcolor{yellow!30}0.5186 \\
      \cellcolor{red!30}irs & \cellcolor{red!30}federal reserve & \cellcolor{red!30}0.5181 \\
      \cellcolor{yellow!30}ds & \cellcolor{yellow!30}puppets & \cellcolor{yellow!30}0.5050 \\
      \cellcolor{red!30}gold standard & \cellcolor{red!30}fiat & \cellcolor{red!30}0.5033 \\
      \cellcolor{yellow!30}reptilians & \cellcolor{yellow!30}demons & \cellcolor{yellow!30}0.4839 \\
      \cellcolor{yellow!30}pizzagate & \cellcolor{yellow!30}pedowood & \cellcolor{yellow!30}0.4800 \\
      \hline
    \end{tabular}
    \caption{Romana Didulo channel -- QAnon Telegram dataset.}
    \label{tbl:romanadidulo-alignment}
  \end{subtable}
  \caption{Alignment score of selected shared words. Shared words color code categories: 1)~Red is Sovereign Citizens, 2)~Yellow is Conspiracy Theory, and 3)~Green is COVID-19.}
  \label{tbl:alignment}
\end{table}

\descr{Alignment Measurement Results.}
Our analysis focuses on analyzing the misalignment of words or terms.
Table~\ref{tbl:alignment} shows the selected word pairs with their alignment and similarity scores.
We colored the rows based on three categories: red for Sovereign Citizens, yellow for Conspiracy Theory, and green for COVID-19.

In the Sovereign Citizen category on Bobby Lawrence's channel, the misaligned terms \emph{natural law} and \emph{common law} (see Tables~\ref{tbl:bobbylawrence-1776-alignment}) are often used interchangeably to distinguish them from government-created law. 
Natural law~\cite{Murphy_2019} is rooted in moral theory, while common law~\cite{Hill_2023} evolved from traditional unwritten law in England.

Another misaligned pair is \emph{fiat} and \emph{cryptocurrency}, but contextually, they both refer to currency. 
Fiat money is a government-issued currency not backed by reserves like gold~\cite{FiatMoneyOxfordReference2023}, while cryptocurrency is digital money used for online transactions without a central bank~\cite{papadamou2023hodl}.

In the Sovereign Citizens context, the misaligned terms \emph{corporations} and \emph{government} (or \emph{governments}) are often conflated, with the government referred to as a corporation~\cite{Berger2016,Sarstechi_2020a}. 
Similarly, the misalignment of \emph{bankrupt} and \emph{corporation} suggests the idea that the government or corporation went bankrupt after abandoning the gold standard in 1933~\cite{Berger2016}.

In the Conspiracy Theory category, we find \emph{illuminati} misaligned with \emph{hillary clinton} (see Table~\ref{tbl:bobbylawrence-1776-alignment}). 
However, in the New World Order conspiracy theory context, both terms are related to global elites. 
Other misalignments relate to the New World Order conspiracy theory, like \emph{deep state} and \emph{globalist} (Table~\ref{tbl:bobbylawrence-1776-alignment}) and \emph{cabal} and \emph{globalist} (Table~\ref{tbl:romanadidulo-alignment}).

There is also a misalignment between \emph{reptilians} and \emph{demons}.
Contextually, they both indicate QAnon's adaptation of traditional conspiracy tropes~\cite{Lewis2005,Time_David_Icke_2008} to the current socio-political climate. 
The misalignment of \emph{pizzagate}~\cite{Pizzagate2020} and \emph{pedowood}~\cite{Pedowood2022} shows two different names related to conspiracy theories about child sex trafficking promoted by QAnon.

In the COVID-19 category, \emph{fauci} and \emph{bill gates} are misaligned, but they are aligned in the context of the COVID-19 conspiracy theory involving both individuals~\cite{Huddleston_2021}. 
Additionally, \emph{vax} misaligns with \emph{covid}, as the slang word for vaccination, is closely related to COVID-19. 
Similarly, \emph{protesters} misalign with \emph{rioters}, as protesters are understood as rioters.

\begin{table}[t!]
  \small
  \centering
  \begin{subtable}[t]{0.5\textwidth}
    \centering
    \begin{tabular}{llr}
      \hline
      \textbf{Bobby Lawrence} & \textbf{QAnon} & \textbf{Alignment} \\
      \hline
      \cellcolor{red!30}united state inc & \cellcolor{red!30}bankrupt & \cellcolor{red!30}0.4590 \\
      \cellcolor{red!30}admiralty maritime & \cellcolor{red!30}tribunal & \cellcolor{red!30}0.4156 \\
      \cellcolor{red!30}cqv trust & \cellcolor{red!30}wealth & \cellcolor{red!30}0.3800 \\
      \cellcolor{red!30}sovereigns & \cellcolor{red!30}illegal immigrants & \cellcolor{red!30}0.3540 \\
      \hline
    \end{tabular}
    \caption{Bobby Lawrence channel -- QAnon Telegram dataset.}
    \label{tbl:bobbylawrence-homomorphism}
  \end{subtable}

  \hfill

  \begin{subtable}[t]{0.5\textwidth}
    \centering
    \begin{tabular}{llr}
      \hline
      \textbf{Romana Didulo} & \textbf{QAnon} & \textbf{Alignment} \\
      \hline
      \cellcolor{green!30}bioweapon & \cellcolor{green!30}vaccines & \cellcolor{green!30}0.5760 \\
      \cellcolor{green!30}draconian measures & \cellcolor{green!30}lockdown & \cellcolor{green!30}0.5299 \\
      \cellcolor{green!30}truckers & \cellcolor{green!30}rioters & \cellcolor{green!30}0.5222 \\ 
      \cellcolor{red!30}gold backed & \cellcolor{red!30}fiat & \cellcolor{red!30}0.4909 \\ 
      \cellcolor{yellow!30}khazarian mafia & \cellcolor{yellow!30}jews & \cellcolor{yellow!30}0.4699 \\
      \cellcolor{yellow!30}reptiles & \cellcolor{yellow!30}illuminati & \cellcolor{yellow!30}0.4199 \\ 
      \cellcolor{green!30}genocidal & \cellcolor{green!30}scamdemic &\cellcolor{green!30} 0.4023 \\
      \cellcolor{red!30}traveling & \cellcolor{red!30}visiting & \cellcolor{red!30}0.3967 \\
      \cellcolor{green!30}jabbing & \cellcolor{green!30}killing & \cellcolor{green!30}0.3953 \\
      \hline
    \end{tabular}
    \caption{Romana Didulo channel -- QAnon Telegram dataset}
    \label{tbl:romanadidulo-homomorphism}
  \end{subtable}
  \caption{Conceptual homomorphism alignment score. Shared words color code categories: 1)~Red is Sovereign Citizens, 2)~Yellow is Conspiracy Theory, and 3)~Green is COVID-19.}
  \label{tbl:conceptual-homomorphism}
\end{table}

\descr{Conceptual Homomorphism Results.}
We present selected conceptual homomorphism word or term pairs on Table~\ref{tbl:conceptual-homomorphism}, showing the translation of words or terms in the left dataset that are absent in the right dataset (i.e., translation of words in Bobby Lawrence's and Romana Didulo's channels to the QAnon Telegram dataset).
The rows are color-coded into three categories: red for Sovereign Citizens, yellow for Conspiracy Theory, and green for COVID-19.

In the Sovereign Citizens category, the term \emph{united state inc} is absent in the QAnon Telegram dataset, but its closest image is the term \emph{bankrupt} (see Table~\ref{tbl:bobbylawrence-homomorphism}).
Next pair is the \emph{admiralty maritime} and \emph{tribunal}, where the admiralty or maritime court is often called the tribunal court.
A more specific Sovereign Citizens' jargon is the use of the term \emph{traveling} to refer to driving, believing the idea that ``free'' men and women are not bound by relevant law because they are not transporting commercial goods or paying passengers~\cite{splc_2021}.
This term is absent in the QAnon Telegram dataset, but it is translated \emph{visiting}.

In the Conspiracy Theory category, the term \emph{khazarian mafia} links to the Khazar hypothesis, postulating that Ashkenazi Jews were mainly descended from Khazars~\cite{Harke_2004,Zannettou_Finkelstein_Bradlyn_Blackburn_2020}, finds the conceptual homomorphic with the word \emph{jews} logical.
The pair of words, \emph{reptiles} and \emph{illuminati}, is associated with conspiracy theories about world domination~\cite{Lewis2005,Time_David_Icke_2008,American_Jewish_Committee_2023c,Kelly2023}.

In the COVID-19 category, \emph{bioweapon} is translated to \emph{vaccine}, aligning with the conspiracy theory suggesting the COVID-19 vaccine is a bioweapon~\cite{Reuters_2021b}. 
Similarly, \emph{draconian measure} nearest image \emph{lockdown}, reflect the belief that the COVID-19 lockdown measures are excessive and contribute to genocidal events~\cite{AFP_Canada_2022}. 
The translation between \emph{genocidal} and \emph{scamdemic} captures the conspiracy view of the COVID-19 pandemic. 
Finally, \emph{jabbing} is translated to \emph{killing}, reflecting the conspiracy perspective that vaccination leads to harm.

\descr{Key Takeaways.}
The worldview alignment framework successfully translated words in Bobby Lawrence's and Romana Didulo's channels to their nearest images in the QAnon Telegram dataset from a list of shared words, indicating a fair amount of understanding of beliefs.
The intertwined beliefs demonstrate that the beliefs of both groups (Sovereign Citizens and the QAnon) are either merging or complementing each other.

\section{Discussion \& Conclusion}
This study explored Sovereign Citizens communities on Telegram. 
To shed light on their online activities, we collected data from Telegram channels associated with two Sovereign Citizens individuals, Bobby Lawrence and Romana Didulo. 
Our dataset includes over 81K text messages from Lawrence's channel and more than 807K from Didulo's. 
Our findings highlight distinct focuses within these channels: Lawrence emphasizes education on achieving American State National status and restoring the Constitutional Republic, while Didulo's channel explores current events from unique perspectives and her royal decrees. 
Notably, both channels exhibit lower toxicity levels compared to the QAnon Telegram dataset used for comparison. 
Moreover, our analysis uncovers vocabulary differences between Sovereign Citizens channels and the QAnon dataset. 
However, a closer look reveals shared conceptual themes, suggesting some ideological overlap between Sovereign Citizens and QAnon beliefs.

\subsection{Limitations}
Sovereign Citizens' presence in social media and online communities remains largely unexplored, raising concerns about the interpretation of our dataset. 
While our findings generally align with what little existing literature there is, we reveal an expansion in their activities beyond paper terrorism, an area with limited theoretical groundwork for interpretation.

It is also important to acknowledge that our analysis relies on the accuracy of third-party tools and libraries. 
Thus, our results are constrained by the limitations of these tools. 
For example, Stanza may exhibit inaccuracies and biases in recognizing named entities~\cite{Vajjala2022,Shan2023}, Perspective API in toxicity analysis, and Top2Vec may generate more outliers than anticipated~\cite{Egger2022}. 
However, given the consistency with existing literature, we have confidence in our main findings.

Furthermore, our dataset notably lacks representation from other Sovereign Citizens communities. 
This is partly due to the difficulty in identifying such communities online, exacerbated by the lack of prior research and explicit indicators of Sovereign Citizens affiliation.

\subsection{Implications \& future work}
Our work highlights the urgent need for the research community to broaden its study of online extremism. 
While past studies have mainly focused on far-right and QAnon, our findings uncover a rising presence of Sovereign Citizens online. 
While previous research on Sovereign Citizens has centered on legal issues like paper terrorism, our study reveals a broader range of behaviors beyond just legal challenges.

Our findings demonstrate Sovereign Citizens' persistent efforts to delegitimize the government, posing significant threats to national democracy and societal stability. 
These efforts could fuel the rise of separatist movements or dormant extremist cells, worsening domestic issues. 
Additionally, our research suggests that the Sovereign Citizens movement has merged with extremist ideologies like QAnon, adapting its traits worldwide. 
We have identified potential Sovereign Citizen channels in the U.K. and Australia, highlighting the need for a keen monitoring on social media to address future security challenges. 
A further investigation into Sovereign Citizens across various online platforms is crucial for developing effective moderation strategies.

Furthermore, we recognize the risks associated with studying extremist groups, both for researchers~\cite{doerflerProfessorWhichIsn2021} and the potential misuse of findings~\cite{yudhoatmojoUnderstandingUseEPrints2023}.  
However, we believe the positive impact on researchers, social media platforms, and policymakers outweighs these potential negatives.

\descr{Acknowledgements}
Satrio Yudhoatmojo would like to thank DIKTI-funded Fulbright Grants for supporting his doctoral study at Binghamton University.
This material is based upon work supported by the National Science Foundation under Grant No. IIS-2046590.
\small
\bibliographystyle{abbrv}

%
%
%

%

\end{document}